\documentstyle[12pt,citation]{article}

\font\tenrm=cmr10
\font\tenit=cmti10
\font\elevenbf=cmbx10 scaled\magstep 1
\font\elevenrm=cmr10 scaled\magstep 1
 1

\font\ninerm=cmr9

\def\overlay#1#2{\ifmmode%
\setbox0=\hbox{$#1$}%
\setbox1=\hbox to\wd0{\hss$#2$\hss}\else%
\setbox0=\hbox{#1}%
\setbox1=\hbox to\wd0{\hss#2\hss}\fi%
 #1\hskip-\wd0\box1 }

\renewenvironment{thebibliography}[1]
 { \elevenrm
   \begin{list}{\arabic{enumi}.}
    {\usecounter{enumi} \setlength{\parsep}{0pt}
     \setlength{\itemsep}{3pt} \settowidth{\labelwidth}{#1.}
     \sloppy
    }}{\end{list}}

\begin{document}
\begin{center}{{\bf Strong $W_L W_L$ Scattering at Hadronic Supercolliders
\footnote{\ninerm Talk given at the ``Workshop on Physics and Experiments with
Linear $e^+e^-$ Collider", Waikoloa, Kona, Hawaii}
\\}
\vglue 5pt
{\rm Kingman Cheung\\}
\baselineskip=13pt
{\tenit Department of Physics, Northwestern University, \\}
\baselineskip=12pt
{\tenit Evantson, Illinois  60208, U.S.A.\\}
\vglue 0.3cm
{\tenrm ABSTRACT}}
\end{center}
\vglue 0.3cm
{\rightskip=3pc
 \leftskip=3pc
 \tenrm\baselineskip=12pt
 \noindent
We will summarize some aspects of the scenario of a
strongly interacting symmetry breaking sector via which the
longitudinal vector boson scattering becomes strong.
We will examine the feasibility of observing  such strong $W_L W_L$
signal at the future  hadronic supercolliders.
\vglue 0.6cm}
\baselineskip=14pt

So far very little is known about the symmetry-breaking-sector (SBS),
except it gives  masses to $W$ and $Z$ bosons, which are  in good agreement
 with experiments.
If no Higgs boson is found below 800 GeV, the heavy Higgs scenario
($\approx 1$ TeV) will imply a strongly interacting SBS
because  the self-coupling $\lambda\sim m_H^2$ becomes strong.
However, there is no evidence to favor the models with  scalar Higgs bosons.
On the  other hand, the standard model without a Higgs boson cannot be a
complete theory as unitarity  breaks down at
1--2 TeV scale, so some kind of new physics must come in to
play.  There have been a number of possible models proposed for the SBS, {\it
e.g.}, technicolor models, which can only be examined in the
future hadronic supercolliders.

One of the best way to uncover the dynamics of the SBS is to study the
longitudinal vector boson scattering.  Equivalence Theorem \cite{ET} states
that at
very high energy the longitudinal part of the vector bosons behave as the
corresponding Goldstone bosons (GB).  These GB's originate from the SBS so that
their scattering must be via the interactions of the SBS, and therefore
 can reveal the  dynamics of the SBS.

One subtle problem is to define what the signal and background are.  We take
the signal to be the process $pp\rightarrow V_L V_LX$, where $L$ refers to
the longitudinal polarization.  Here the $V$ refers to $W$ and $Z$ bosons.
It is clear that only the longitudinal part is
sensitive to new physics of the SBS.  The most obvious irreducible background
is the electroweak productions (EW) of $pp\rightarrow V_T V_T X,\, V_L V_T X$,
 where $T$ stands for transverse polarization.  We will use the SM with a
light Higgs ($m_H=0.1$~TeV) to represent this background.
Other backgrounds in
general include the lowest order production of vector boson pairs $VV$ plus
higher order QCD jet emissions, and  heavy quark production in which the
heavy quark decays into leptons.

Different vector boson pair channels have
different backgrounds, which depend on the decay modes of the vector boson
pair.  In the following we consider only the ``gold-plated'' mode,
\begin{equation}
W\rightarrow \ell\nu,\, \quad Z\rightarrow \ell^+\ell^-\,.
\end{equation}
The backgrounds for each channel are summarized in Table~\ref{table1}.  In
this gold-plated mode we only tag on the leptonic decay of the vector boson
pair, so  all the jet activities come from the spectator quarks.  One of
the advantages  of this mode is that the spectator quarks in the signal
events are  very different from  the jets associated with the backgrounds.
We can then make use of these spectator quarks to come up with the techniques
of forward jet-tag \cite{ZZ} and central jet-veto \cite{W+W+}
 to suppress backgrounds.  The hadronic decay
mode of the boson pair will dilute the distinct characteristics of the
spectator jets in the gold-plated mode.  Though the hadronic mode has a larger
branching ratio it suffers from much larger background.  More studies in
this mode  are needed.  We will list a few models in the next section, and
present the strategies used in the $V_L V_L$ scattering in sec.~2, and show
how the backgrounds are suppressed in $W^+W^-$ channel in sec.~3.
Finally, the predictions for the models will be presented  in sec.~4.
\begin{table}[t]
\caption{\label{table1}
\tenrm \baselineskip=12pt
Table summarizing various backgrounds present in each vector boson
channel when the gold-plated mode is considered.}
\baselineskip=14pt
\begin{center}
\begin{tabular}{|c|c|c|}
\hline
Channel    &   signal    & backgrounds  \\
\hline
$ZZ$      &   $Z_LZ_L \rightarrow \ell^+\ell^-\ell^+\ell^-$  & $Z_T Z_T$, QCD
$ZZ+$ jets \\
$W^+W^-$  & $W^+_LW^-_L \rightarrow \ell^+\nu \ell^-\bar\nu$ & $W^+_T W^-_T$,
QCD $W^+W^- +$ jets, $t\bar t(g)$ \\
$W^+W^+$  & $W^+_L W^+_L \rightarrow \ell^+\nu \ell^+\nu$ &  $W^+_T W^+_T$,
$W^+W^+$(gluon-exch.), $W^+t\bar t$  \\
$W^+Z$   & $W^+_L Z_L \rightarrow \ell^+\nu \ell^+\ell^-$ &  $W^+_T Z_T$, QCD
$W^+Z +$ jets, $Zt\bar t(g)$ \\
\hline
\end{tabular}
\end{center}
\end{table}
\vglue 0.6cm
{\elevenbf \noindent 1. Models for SBS}
\vglue 0.4cm

Different type of models have been proposed for SBS. In the following we
will list a few of them.  A lot of these models are based on Low Energy
Theorem \cite{LET}(LET), which tells the low energy behavior of the scattering
amplitudes.  The amplitudes derived from different models should become
LET-derived amplitudes  at low energy limit because it is required  by
the symmetry.
When these models are extrapolated to high energy unitarity has to be
conserved, or some unitarization schemes are needed.

Models can be classified according to the spin and isospin of the resonance
fields. \\
(i) Scaler-like models: these models are motivated if we think of the GB's in
terms of the linear $\sigma$ model we expect the $V_L V_L$
scattering amplitudes  ${\cal M}$ at high energy
will be unitarized by some  spin-0 and isospin-0 scalars.
SM with a heavy Higgs is an obvious example.  Other examples include $O(2N)$
model \cite{einhorn}  and the model with chirally coupled scalar fields
\cite{scalar}. \\
(ii) Vector-like models: these models are motivated if we think of the GB's in
analogy with the pions of the hadronic QCD associated with the light quarks
we expect the ${\cal M}$ will be unitarized by
some spin-1, isospin-1 vector resonances.  Typical examples are the BESS
model \cite{scalar} and technirho \cite{bigpaper}. \\
(iii) Nonresonant models:  they are the  models for describing
the $V_L V_L$ scattering which becomes   strong
below the threshold  for resonance production.  In these models  no
resonance is
present at the scale of $V_LV_L$ scattering so explicit unitarization is
prescribed to preserve unitarity.  Typical examples are the one considered by
Chanowitz and Golden \cite{chan},
in which the LET amplitudes are allowed to grow with $s$
until the partial wave coefficients saturate the bound $|a^I_\ell|<1$, and
models with $K$-matrix unitarizations \cite{bigpaper}.

\vglue 0.6cm
{\elevenbf \noindent 2. Strategies in Studying $W_LW_L$ Scattering}
\vglue 0.4cm

So far there is no systematic way to incorporate these models with the rest of
the SM exactly, we have to rely on the
Effective W Boson Approximation \cite{EWA} (EWA), which
gives the effective vector-boson luminosities coming off the quarks.
But in EWA kinematics of the spectator quarks are not included exactly.
Only the SM with a heavy Higgs boson can be computed exactly with the correct
dynamics of the scattered quarks.  We will use the SM with a heavy Higgs boson
$(m_H=1$~TeV) to differentiate   against the backgrounds, and
to come up with the jet-tagging
and jet-vetoing efficiencies, and apply these efficiencies to the other
models.  This is justified as the dynamics of the spectator quarks should be
independent of the $V_L V_L$ scattering.  Next step is to differentiate the
characteristics of the signal and various backgrounds.

The leptons decaying from the vector boson pair for the signal are quite
different from those of the backgrounds.  This is due to the fact that at high
energy the interaction between $V_LV_L$ becomes strong under the
scenario of a strongly interacting SBS.  We expect the $V_LV_L$ scattering to
be enhanced  at high invariant mass region.  For the EW, QCD and heavy quark
backgrounds no enhancement is likely at high invariant mass region.
Therefore,  in the signal the scattered vector boson pair has a relatively
large invariant mass and energy, implying the leptons decaying from them are
also very energetic and have large $p_T$.
Other features of these leptons are that
they are isolated and very back-to-back \cite{gunion}, due to the
fact that the $V_LV_L$ coming off the strong scattering are very  back-to-back
and energetic.  Since there is no longitudinal boost in the transverse plane,
so the back-to-back feature is most transparent in the transverse plane.
We can come up with a set of cuts ($e.g.\;p_{T\ell},\,|y_\ell|, \Delta
p_{T\ell\ell}$) on the leptons  to reduce the backgrounds.  Besides, isolation
cuts on the charged leptons are required to
 eliminate the reducible backgrounds from the
semileptonic decays of the $b$ and $c$ quarks \cite{gunion}.

It is important to note that two spectator quarks always emerge in
association with the $V_L V_L$ scattering signal.  The incoming quark with
energy of order 1 TeV radiates a $V_L$, which is not much off-shell.  The
recoiled quark then tends to carry most of the energy of the incoming one
and so  it is also energetic of order 1 TeV.  By helicity argument
the quark  after radiating a longitudinal vector boson
is favored in the forward direction.
We therefore expect by tagging energetic ($O(1$ TeV)) jets  in the
forward/backward rapidity regions we can enhance the signal:background ratio.
In particular, the continuum pair productions do not
have any QCD jets at lowest order, and additional gluon radiations
are relatively soft.
It has been recently suggested that
tagging just one of the spectator quarks as a single energetic jet
can be just as efficient in suppressing the backgrounds,
and far more efficient in retaining the signal \cite{ZZ}.
We will apply the forward jet-tag to most of the channels
that we are considering.

\begin{table}[t]
\caption{\label{table2}
\tenrm \baselineskip=12pt
Table summarizing the cuts, jet-tag and jet-veto efficiencies
in $ZZ$, $W^+W^-$, $W^+W^+$, and $W^+Z$ channels in the studies of strong
$V_L V_L$  scattering at SSC (LHC).}
\baselineskip=14pt
\begin{center}
\begin{tabular}{|c|c|}
\hline
 $ZZ$           &           $W^+W^-$ \\
\hline
$|y_\ell| < 2.5$            &            $|y_\ell| <2$ \\
$p_{T\ell}>40$ GeV          &            $p_{T\ell}>100$ GeV \\
$p_{TZ} > \frac{1}{4}\sqrt{m^2_{ZZ}-4m_Z^2}$ & $\Delta p_{T\ell\ell}>450$ GeV
\\
$m_{ZZ}>500$ GeV            &                 $\cos \phi_{\ell\ell} <-0.8$ \\
                            &                 $ m_{\ell\ell}>250$ GeV \\
tag: $E_j({\rm tag})>1(0.8)$ TeV  &      tag: $E_j({\rm tag})>1.5(1.0)$ TeV \\
\mbox{\hspace{0.3in}} $3<|\eta_j({\rm tag})| < 5$  &
\mbox{\hspace{0.3in}} $3<|\eta_j({\rm tag})| < 5$ \\
                    & veto: $p_{Tj}({\rm veto}) > 30$ GeV \\
                    & \mbox{\hspace{0.3in}} $|\eta_j({\rm veto})| < 3$ \\
tag eff.: 59(49)\%          &   veto eff.: 57(40)\% \\
                           &   veto+tag eff.: 38(24) \% \\
\hline
 $W^+W^+$           &           $W^+Z$ \\
\hline
$|y_\ell| < 2$            &            $|y_\ell| <2.5$ \\
$p_{T\ell}>100$ GeV          &            $p_{T\ell}>40$ GeV \\
$\Delta p_{T\ell\ell}>200$ GeV &        $\overlay{/}{p}_T>75$ GeV \\
$\cos \phi_{\ell\ell} <-0.8$    &        $ p_{TZ} > \frac{1}{4} m_T$ \\
$ m_{\ell\ell}>250$ GeV        &        $m_T > 500$ GeV \\
                              &      tag: $E_j({\rm tag})>2(1.5)$ TeV \\
                & \mbox{\hspace{0.3in}} $3<|\eta_j({\rm tag})| < 5$ \\
veto: $p_{Tj}({\rm veto}) > 60$ GeV & veto: $p_{Tj}({\rm veto}) > 60$ GeV \\
\mbox{\hspace{0.3in}} $|\eta_j({\rm veto})| < 3$ &
 \mbox{\hspace{0.3in}} $|\eta_j({\rm veto})| < 3$ \\
veto eff.: 69(58)\%       &   veto eff.: 75(48)\% \\
                         & veto+tag eff.: 40(20)\% \\
\hline
\end{tabular}
\end{center}
\end{table}

Another crucial point to note is that the initial $V_L$'s
participating in the $V_L V_L$ scattering have a $1/(p_T^2+m_V^2)^2$
distribution with respect to the quarks from which they are emitted.
This is to be contrasted, for instance, with $V_T V_T$ scattering where
the initiating $V_T$'s have a $p_T^2/(p_T^2+m_V^2)^2$ distribution with
respect to the emitting quarks.  We therefore  expect the quarks in the
$V_LV_L$ case has a softer $p_T$ spectrum, and, as mentioned above, preferably
 in the
forward/backward region.  We can then impose a central jet-veto to suppress
the $V_TV_T$ background \cite{W+W+,gunion}.
Besides, the heavy quark productions like the $t\bar
t$ will give rise to a lot of central jet activities after decaying into $b$'s.
The central jet-veto is  very effective in suppressing these heavy quark
backgrounds as well \cite{WW}.

We will show how the cuts considered
 here are applied to the $W^+W^-$ channel to
get rid of the backgrounds.  This channel is the most challenging due to the
huge $t\bar t$ background.  As space is limited the cuts used in the
other channels  are summarized in Table~\ref{table2}.

\vglue 0.6cm
{\elevenbf \noindent  3. $W^+W^-$ Channel}
\vglue 0.4cm
\begin{figure}[t]
\vspace{3in}
\caption{\label{veto}
\tenrm \baselineskip=12pt
(a) The pseudorapidity distribution of the veto-jet candidate, (b)
the energy distribution of the tagged jet in the forward rapidity region,
3$<|\eta_j({\rm tag})|<$5.}
\end{figure}

The $W^+W^-$ channel for the SM with a heavy higgs was studied in
Ref.~11.  The background includes QCD production of $W^+W^- + $ jets,
EW background, and $t\bar t$.  The crucial cut used is the central jet-veto.
The $b$'s  decaying from the $t\bar t$  give rise to a lot of central jet
activities.  We then impose a veto of
\begin{equation}
\label{jetveto}
p_{Tj}({\rm veto}) > 30\;{\rm GeV} \quad {\rm and}\quad
|\eta_j({\rm veto})|<3\,,
\end{equation}
for both SSC and LHC energies.
The pseudorapidity of the veto-jet candidate is shown in Fig.~\ref{veto}(a).
All the events on the curves below $|\eta_j({\rm veto})|<3$ are rejected by the
veto of Eq.~(\ref{jetveto}).  What remain are those events above $|\eta_j({\rm
veto})|=3$ and those events with $p_{Tj} < 30$ ~GeV.
It reduces the $t\bar t$ background by about two order of magnitudes
while only cuts less than half on the signal.
This veto is even more  effective for a heavier top because the heavier top
 gives larger average $p_T$ to the $b$.
We also have to use forward jet-tag
 to further reduce the $t\bar t$ background and the QCD $WWj$
background. The energy distributions of the tagged jet are
 shown for SSC in Fig.~\ref{veto}(b).   We choose a forward jet-tag of
\begin{equation}
\label{jettag}
E_j({\rm tag})>1.5 (1.0)\;{\rm TeV} \quad {\rm and} \quad 3<|\eta_j({\rm
tag})|<5\,,
\end{equation}
for SSC (LHC).
\begin{figure}[t]
\vspace{3in}
\caption{\label{dptll}
\tenrm \baselineskip=12pt
The $\Delta p_{T\ell\ell}$ distributions for the SM with a heavy Higgs
boson ($m_H$=1 TeV), and EW, QCD and $t\bar t$ backgrounds.}
\end{figure}
\baselineskip=14pt
Besides, we impose rather stringent cuts on the charged lepton pair
\begin{equation}
\label{leptoncut}
p_{T\ell}>100\;{\rm GeV},\quad |y_\ell|<2,\quad \Delta p_{T\ell\ell}>450\;{\rm
GeV}\,,
\end{equation}
where $\Delta p_{T\ell\ell}$ is defined as $|\vec p_{T\ell_1} - \vec
p_{T\ell_2}|$, the vector difference between the
transverse momenta  of the two charged leptons.  As mentioned in the last
section, the leptons from the $V_LV_L$ scattering are very back-to-back. The
dependence of the differential cross section on $\Delta p_{T\ell\ell}$ is
shown in Fig.~\ref{dptll}.

\begin{table}[t]
\caption{\label{table3}
\tenrm \baselineskip=12pt
Number of events observed in one SSC (LHC) year, assuming integrated
luminosities of 10 (100) fb$^{-1}$.  The acceptance cuts for each channel is
shown in Table 2.}
\baselineskip=14pt
\begin{center}
\begin{tabular}{|@{\extracolsep{0.15in}}c||cccc|}
\hline
        & $ZZ$      &     $W^+W^-$     &     $W^+W^+$   &  $W^+Z$   \\
\hline
bkgd  &    1.0 (1.0)   &    21 (18)   &   3.5 (6.2)   &   2.5 (2.4) \\
\hline
SM    &    11 (14)   &    48 (40)   &   6.4 (9.6)   &   1.3 (1.0) \\
scalar&    6.2 (7.5)    &    30 (26)   &   8.2 (12)   &   1.8 (1.4) \\
O(2N)&    5.2 (6.4)    &    24 (19)   &   7.1 (10)   &   1.5 (1.1) \\
\hline
Vec 2&    1.1 (1.4)    &    15 (8.0)   &   7.8 (12)  &   9.5 (4.8) \\
Vec 2.5&    1.5 (1.7)    &   12 (6.8)   &   11 (16)  &   6.2 (3.2) \\
\hline
LET CG &    2.6 (2.5)    &   16 (9.2)   &   25 (27)  &   5.8 (3.2) \\
LET K &    2.2 (2.2)    &   12 (7.2)   &   21 (24)  &    4.9 (2.9) \\
\hline
\end{tabular}
\end{center}
\end{table}

\vglue 0.6cm
{\elevenbf \noindent  4. Predictions}
\vglue 0.4cm

Here the results are presented with the leading order parton distribution of
Morton and Tung \cite{MT} with the scale $Q^2=m_W^2$ for the signal.  The
number of signal events predicted  by each model for SSC and LHC are shown in
Table~\ref{table3}, assuming integrated luminosities of 10 and 100 fb$^{-1}$
respectively.  We can see from Table~\ref{table3}
 that different channels are sensitive
to different types of new physics.
The $W^+W^-$ and $ZZ$ channels have the largest
signal:background ratio for the scalar-like models.  Vector-like  models are
more likely to be discovered in the $WZ$ channel, and $W^+W^+$ are more
sensitive to nonresonant models.  Another informative quantity is the number
of SSC/LHC years needed to discover a particular model in each channel.
The observability is defined as when the maximum number of background events
$B_{\rm max}$ at a certain
confidence level is less than the minimum number of signal
plus background events $(S+B)_{\rm min}$ at the same confidence level, where
the numbers $B_{\rm max}$ and $(S+B)_{\rm min}$ are computed using Poisson
statistics.  Table~\ref{table4} shows the number of SSC(LHC) year needed to
discover a particular model for each  channel at 99\% confidence level.
\begin{table}[t]
\caption{\label{table4}
\tenrm \baselineskip=12pt
Number of SSC (LHC) years needed to discover the models at 99\% confidence
level.  The  entry is empty if it is greater than 10 years.}
\baselineskip=14pt
\begin{center}
\begin{tabular}{|@{\extracolsep{0.15in}}c||cccc|}
\hline
        & $ZZ$      &     $W^+W^-$     &     $W^+W^+$   &  $W^+Z$   \\
\hline
SM    &    2.2 (2.0)    &    0.5 (0.75)   &   6.2 (5.2)   &   - (-) \\
scalar&    4.0 (3.0)     &    1.0 (1.0)    &   4.0 (3.2)   & - (-)  \\
O(2N)&     5.8 (4.8)   &    1.5 (1.5)   &   4.5 (4.2)   &   - (-)  \\
\hline
Vec 2&     - (-)  &    2.2 (5.0)   &   4.8 (3.5)   &   1.5 (3.0)  \\
Vec 2.5&   - (-)  &    3.5 (7.0)   &   2.2 (2.0)   &   2.8 (6.8) \\
\hline
LET CG &   7.8 (9.0)    &   2.5 (4.5)  &   0.5 (0.75)   &   3.2 (7.8) \\
LET K &    - (-)   &    4.0 (7.0)  &  0.75 (0.75)  &  4.2 (9.5) \\
\hline
\end{tabular}
\end{center}
\end{table}

In conclusions, within 2--3 years of running at SSC/LHC we can either discover
 or exclude the possibility of a strongly interacting SBS via the longitudinal
vector boson scattering in the gold-plated decay mode.  Even though in the
near future a light Higgs boson is found the studies of longitudinal vector
boson scattering are still worthwhile  to make sure that the bad high
energy behavior of $V_L V_L$ scattering
is cured, or some  new physics must come in at TeV scale.

\baselineskip=14pt

\vglue 0.6cm
{\elevenbf \noindent Acknowledgements:}
Thanks to J.~Bagger, V.~Barger, J.~Gunion, T.~Han, G.~Ladinsky,
R.~Rosenfeld, C.~P.~Yuan, and  D.~Zeppenfeld for collaboration
on different parts of the work  presented here.
Special thanks to the organizers of the workshop for financial support,
and to S.~Kanda and G.~Hou for their  hospitality during the workshop.
This work was supported in part by the DOE grant DE-FG02-91-ER40684.

\vglue 0.6cm
{\elevenbf \noindent Reference}
\vglue 0.4cm

\end{document}